\documentstyle[aps,prl]{revtex}

\begin{document}
\pagestyle{empty}
\narrowtext
\twocolumn

\noindent
{\bf Hopf Term for a Two-Dimensional Electron Gas} \\

In a recent Letter \cite{ApelBychkov}, Apel and Bychkov presented
microscopic calculations of the prefactor $\Theta$ of the topological
Hopf term in the effective action of a two-dimensional electron gas in
a magnetic field at the filling factor $\nu=1$. They suggested that
``this is the first case in condensed matter theory in which one can
calculate a nonzero Hopf term from the microscopic model.''

First, this is not true.  A nonzero $\Theta$ was calculated
microscopically for $^3$He-A films in Ref.\ \cite{VolovikYakovenko},
and the results were summarized in book \cite{VolovikBook}.  The
method of Ref.\ \cite{VolovikYakovenko} was used to microscopically
calculate $\Theta$ for various lattice models in Ref.\
\cite{Yakovenko1990} and for the magnetic-field-induced
spin-density-wave (FISDW) in quasi-one-dimensional organic conductors
in Ref.\ \cite{Yakovenko1991}. The results of Ref.\
\cite{Yakovenko1990} were summarized in conference proceedings
\cite{Yakovenko1989}.

Second, the derivation of the Hopf term in Ref.\ \cite{ApelBychkov} is
not rigorous enough and cannot be considered as a proof. The
parametrization of the rotation matrices and the unit vector $\vec n$
in Ref.\ \cite{ApelBychkov} by the Euler angles $\bar\theta$,
$\bar\phi$, and $\bar\psi$ is potentially dangerous, because the
angles $\bar\phi$ and $\bar\psi$ are ill-defined where
$\cos{\bar\theta}=\pm 1$.  Moreover, the integrand in Eq.\ (14) of
Ref.\ \cite{ApelBychkov} is the total spatial derivative:
\begin{equation}
-\partial_x\left[\cos\bar\theta~{\partial(\bar\phi,\bar\psi) \over
\partial(t,y)}\right] +
\partial_y\left[\cos\bar\theta~{\partial(\bar\phi,\bar\psi) \over
\partial(t,x)}\right].
\label{1}
\end{equation}
The space integral of Eq.\ (\ref{1}) is exactly zero, if one rotates
the field $\vec{n}$ in such a way that
$\cos\bar\theta(\infty)=0$. Since the Hopf term is invariant under
such rotation, this means that Eq.\ (14) of Ref.\ \cite{ApelBychkov}
does not contain the Hopf term at all.  This zero result is probably
an artifact of their parametrization.  In the previous derivations of
the Hopf term \cite{VolovikYakovenko,Yakovenko1990,Yakovenko1991} the
parametrization in terms of the Euler angles was avoided.

A general class of mean-field fermion models characterized by a
microscopic Hamiltonian of the form
\begin{equation}
\hat{H}=\hat{H}_0+\vec{\sigma}\vec{n}(\vec{r},t)\,\hat{H}_1
\label{H}
\end{equation}
was considered in Ref.\ \cite{Yakovenko1990}. In Hamiltonian (\ref{H}),
which acts on the electron wave functions, $\vec{\sigma}$ are the Pauli
matrices that act on the spin indices of the electrons,
$\vec{n}(\vec{r},t)$ is a unit vector slowly varying in
(2+1)-dimensional space-time, and the spin-independent Hamiltonians
$\hat{H}_0$ and $\hat{H}_1$ are such that the system has an insulating
energy gap at the Fermi level. (In the case of the BCS superconducting
gap \cite{VolovikYakovenko}, the equations below are similar, but
somewhat modified.) As shown in Refs.\ \cite{VolovikYakovenko,Yakovenko1990},
the effective action of model (\ref{H}) (obtained by integrating out the
electrons) contains the Hopf term, whose coefficient $\Theta$ is given
by the following expression (in the normalization of Ref.\
\cite{ApelBychkov}):
\begin{equation}
\Theta=\pi N,
\label{Theta}
\end{equation}
where $N$ is an integer-valued topological invariant in the momentum
space (see also Ref.\ \cite{Wiegmann}):
\begin{equation}
    N=\frac{1}{4\pi^2} {\rm Tr}\int
    d\omega\,dk_x\,dk_y\;G\frac{\partial G^{-1}}{\partial \omega}
    G\frac{\partial G^{-1}}{\partial k_x} G\frac{\partial
    G^{-1}}{\partial k_y}.
\label{N}
\end{equation}
In Eq.\ (\ref{N}), $k_x$ and $k_y$ are the electron wave vectors
in the $x$ and $y$ directions, $\omega$ is the Wick-rotated
frequency, and
\begin{equation}
G(\omega,k_x,k_y)=[i\omega-\hat{H}(k_x,k_y)]^{-1}
\label{G}
\end{equation}
is the Green function of the electrons. The Hamiltonian $\hat{H}$ in
Eq.\ (\ref{G}) is given by Eq.\ (\ref{H}) with the field $\vec{n}$
being uniform in space-time. (To derive the Hopf term, we locally
transform the electrons $\psi'=\hat{U}(\vec{r},t)\psi$ to make
$\vec{n}$ uniform and expand the effective action in the gradients of
$\hat{U}(\vec{r},t)$ \cite{VolovikYakovenko,Yakovenko1990}.) It is
assumed that $k_x$ and $k_y$ are good quantum numbers, thus $\hat{H}$
and $G$ are diagonal in $k_x$ and $k_y$. The topological invariant
(\ref{N}) also determines the quantized Hall conductivity of the
system:
\begin{equation}
\sigma_{xy}=N e^2/h,
\label{xy}
\end{equation}
so $\Theta$ and $\sigma_{xy}$ are proportional to each other.

Thus, for a model (\ref{H}), a microscopic derivation of $\Theta$
amounts to plugging the Green function of the model into Eq.\ (\ref{N})
and doing the integral. Since the mean-field Hartree-Fock model of Ref.\
\cite{ApelBychkov} belongs to the class (\ref{H}), Eqs.\ (\ref{Theta}),
(\ref{N}), and (\ref{xy}) should apply to this model. Comparing the
value of the Hall conductivity at $\nu=1$ with Eq.\ (\ref{xy}), one
finds that $N=1$, thus, from Eq.\ (\ref{Theta}), $\Theta=\pi$, as
suggested in Ref.\ \cite{ApelBychkov}. Strictly speaking, integral
(\ref{N}) has to be somewhat modified for this model, because $k_x$ and
$k_y$ are not good quantum numbers in the magnetic field simultaneously.
That amounts, basically, to replacing the integration over the wave
vectors by averaging over the phases of the boundary conditions, which
is standard in the quantum Hall effect theory. \\

\noindent
G.~E.~Volovik$^{1,2}$ and V.~M.~Yakovenko$^3$

$^1$Helsinki University of Technology, Low Temperature Laboratory,
P. O. Box 2200, FIN-02015 HUT, Finland

$^2$Landau Institute for Theoretical Physics, 117334 Moscow, Russia

$^3$Department of Physics, University of Maryland, College Park, MD
20742-4111, USA \\
\\
March 26, 1997, {\bf cond-mat/9703228} 

\end{document}